\documentclass{article}
\usepackage[utf8]{inputenc}
\usepackage{graphicx}
\usepackage{amsmath}
\usepackage[export]{adjustbox}
\usepackage{dirtytalk}
\usepackage{wrapfig}
\usepackage{color}

\title{Time Evolution of Pulsar Magnetosphere I - An Implicit Approach}
\author{Sushilkumar Sreekumar\footnote{This paper is part of the dissertation presented by Sushilkumar Sreekumar in partial fulfillment of the requirements for the degree of Doctor of Philosophy in Physics at University of Texas San Antonio.} \footnote{Department of Physics and Astronomy, University of Texas San Antonio (UTSA)} \\
Eric M. Schlegel \footnote{Vaughan Family Professor, UTSA}}

\date{April 2018}

\begin{document}

\maketitle

\textbf{Abstract} We apply a
computationally efficient
technique to validate the global structure of the pulsar magnetosphere. In this first of a series of studies, a 3D, computationally intensive, implicit Crank-Nicolson finite-difference scheme is developed. The region of magnetic influence is evolved under the approximation of force$-$free electrodynamics. The main objective of this paper is to present our code and use it to demonstrate and verify the now widely - accepted global features of a pulsar magnetosphere. Our results qualitatively agree with previously developed time-dependent models for an oblique rotator. In line with earlier studies, we also demonstrate that our simulations can run for many stellar rotations. Once we extend our code, we believe that our implicit approach can be extremely useful to investigate magnetospheres filled with resistive plasma, develop better resolution current sheets and investigate small scale microphysics of pair creation using particle-in-cell techniques.

\section{Introduction}

Neutron stars (NSs) are formed in the supernova explosions of massive stars. These remnant sources are extremely small, rapidly rotating, acutely dense and possesses a surface magnetic field strength (up to $10^{14}$ G ) unlike other astrophysical sources \cite{1991tnsm.book.....M}. These unique attributes of NSs provide a test bed to understand the physics in a wide range of areas, especially plasma physics and pulsar magnetospheres. It has been long established \cite{1969ApJ...157..869G} that a NS must support a very strong magnetosphere which corotates at the same angular velocity as the remnant star. The corotating region near the NS surface  is filled with plasma with densities ranging from $10^{11} - 10^{12}$ particles/cm$^3$. These densities are greater than the canonical Goldreich$-$Julian charge density $(\rho_{GJ} = \frac{\bar \Omega \cdot \bar B}{4\pi})$ which is the minimum density required to maintain corotation. Abundant plasma within the magnetosphere can be generated through pair-creation and other acceleration mechanisms. These acceleration mechanism play a critical role as they not only replenish the magnetosphere with particles but can also reproduce observed pulsar light phenomenology.

Understanding of the pulsar magnetosphere over the past 50 years has been carried out along two paths. \textit{Local models} try to mimic the magnetosphere by assuming gaps to occur at specific locations in an otherwise plasma filled magnetosphere. An outer-gap model (OG) \cite{1986ApJ...300..500C},\cite{1995ApJ...438..314R} is developed where gaps are assumed along the outer regions of the magnetosphere. Recent results from the Fermi Gamma-Ray Space Telescope (FGST) supports high-energy emission originating from the outer regions of the magnetosphere \cite{2010ApJS..187..460A} but these OG models underestimate the observed power. The alternatives, polar cap (PC) models occur when gaps are assumed to develop on the polar cap of the NS. A PC model with super exponential spectral cut off \cite{1986ApJ...300..500C} has been developed. However, the high-energy spectral cutoff for the Vela pulsar rules out the PC emission models. Other candidate models include the numerical separatrix model \cite{2010ApJ...715.1282B} and \cite{2010MNRAS.404..767C}, Slot Gap (SG) model \cite{2003ApJ...588..430M}, and striped pulsar wind model \cite{2005ApJ...627L..37P} are other possible candidates. These local models have a more realistic appeal as they are capable of calculating particle acceleration and electromagnetic emissions compared to the traditional vacuum model. However, uncertainties still exist over the exact location, acceleration and emission mechanisms in the magnetosphere.

Nonetheless, local models decouple themselves from the \textit{global} structure of the magnetosphere. This work is focused on exploring the global features (see below) of the magnetosphere; the reader interested in more details of the \textit{local} models should consult the above cited references.

\textit{Global models}, on the other hand, mimic a pulsar magnetosphere by accounting for global features such as the light cylinder (LC) and global current distribution. These models are developed assuming the force-free electrodynamics (FFE) (\cite{1999ApJ...511..351C}, \cite{2006ApJ...648L..51S}) or relativistic magnetohydrodynamics (RMHD, \cite{2006MNRAS.367...19K}). Field dynamic calculations for resistive \cite{2012ApJ...746...60L}, dissipative \cite{2012ApJ...749....2K} and kinetic \cite{2017arXiv171003170K} approaches have also been developed. These global field dynamic calculations have not been included to handle particles (see section $2.2$ for details) or acceleration mechanisms and therefore unlike local models cannot match the observed light phenomenology. An excellent review on the electrodynamics of pulsar magnetosphere and its advances in pulsar modelling can be found in \cite{2017SSRv..207..111C} and  \cite{2017arXiv170200732V} respectively.

Only recently, attempts have been made to develop a self-consistent approach by combining these local and global models using a particle-in-cell (PIC) technique (\cite{2010PASJ...62..131U}, \cite{2014ApJ...795L..22C}, \cite{2015MNRAS.448..606C}, \cite{8209414} and \cite{2015ApJ...801L..19P}). But, these 3D PIC simulations are not fully self-consistent as they are unable to simulate small scale microphysics and hence necessitate hybrid \cite{2017arXiv171003170K} or implicit codes. 

As a result, in this first of a series of studies, a 3D, \textit{implicit}, field dynamic, global structure of the magnetosphere is developed under ideal MHD and FFE condition. This paper is aimed to establish our numerical scheme, code check and its implementation using the FFE approximation. The ultimate goal is, however, to develop an implicit PIC code which is able to resolve small-scale features of a pulsar magnetosphere. Such an \textit{unconditionally stable} code will be able to handle a particle motion solver and electromagnetic field solver.

In section 2, we describe the main motivation for our work and how our numerical approach can be useful in demonstrating a more realistic magnetosphere. Section 3 describes the computational and electrodynamic requirements and illustrates several validation criteria for a magnetosphere to achieve a steady state. In section 4, we present the result of our code check by reproducing the results for a NS surrounded by vacuum and for a magnetosphere completely filled with plasma. A comparison of our results with the current understanding of the structure of the magnetosphere is presented in section 5. In section 6, we describe our short-and-long term planned work for the future. The Appendix includes a recipe to set up one update equation in section 7.

\section{MOTIVATION}

\subsection{\textbf{Vacuum $\rightarrow$ FFE magnetosphere}}

Modelling of a pulsar magnetosphere depends on the amount of plasma present in the magnetosphere. Two extreme cases exist: \textit{vacuum magnetosphere}, where a NS is surrounded by vacuum and the \textit{force-free magnetosphere}, which is a plasma-filled solution. Intermediate to these two extreme cases we have the resistive solution.

The solution to a vacuum dipole magnetosphere was given by Deutsch \cite{1955AnAp...18....1D} and remains a valid solution for an inclined dipole rotator. The net Poynting flux \cite{1969ApJ...157..869G} for such an inclined rotator in vacuum is given below (equation \textbf{(1)}), where $\mu$, $\Omega$ and $\psi$ are magnetic dipole moment, angular velocity, inclination angle between rotation axis and magnetic axis respectively.

\begin{equation}
    L_{vac}=\frac{2}{3}\frac{\mu^2 \Omega^2}{c^3}sin^2\psi
\end{equation}

The vacuum solution will act as a code check for the implicit solver that has been developed. This is done by setting the source terms in the Maxwell equations ($\textbf{2}$ and $\textbf{3}$) to zero. The dipole conductor is then set into rotation and recovers the vacuum solution within one stellar rotation. Refer to section 4 for a detailed explanation.

The other extreme is a force-free electrodynamic (FFE) solution where the magnetosphere has an abundant supply of charges. The solution can again be developed by using the Maxwell equations in special relativity {(equation $\textbf{2}$} and {$\textbf{3}$)} and evolving it under ideal MHD {(equation $\textbf{4}$)} and FFE {(equation $\textbf{5}$)} conditions. Refer to section $3.2$ for the choice of current density.

\begin{equation}
\frac{1}{c} \frac{\partial \bar B}{\partial t} = -\nabla \times \bar E
\end{equation}

\begin{equation}
\frac{1}{c} \frac{\partial \bar E}{\partial t} = -\nabla \times \bar B - \frac{4\pi}{c}\bar J
\end{equation}

\begin{equation}
    \bar E \cdot \bar B = 0
\end{equation}

\begin{equation}
    \rho \bar E + \frac{1}{c}\bar J \times \bar B = 0
\end{equation}

The solution to the FFE problem was first developed numerically by Contopoulous et.al (\cite{1999ApJ...511..351C}, hereafter CKF) for an aligned rotator. The main feature of a CKF-type magnetosphere is the distribution of currents $(I_{CKF})$ along with $\rho_{GJ}$. This solution supports closed field lines which maintain the magnetosphere and open field lines which stretch all the way to infinity. The closed field lines extend outward to the light-cylinder ($R_{LC}=\frac{c}{\omega}$), defining a characteristic length, while open field lines support pulsar winds.
Since then several groups successfully reproduced the CKF-type magnetosphere (\cite{2006ApJ...648L..51S},\cite{2009A&A...496..495K},\cite{2012ApJ...749....2K}) for an inclined rotator. Similar to equation $(1)$, Spitkovsky $(2006)$ \cite{2006ApJ...648L..51S} was able to numerically approximate a simple formula for the net Poynting flux (equation $\textbf{6}$).

\begin{equation}
    L_{vac}\approx \frac{\mu^2 \Omega^4}{c^3}\left (1+sin^2\psi \right )
\end{equation}

The main objective of the present paper is to reproduce the CKF-type magnetosphere field dynamic calculations (section \textbf{$2.2$}) through implicit (section \textbf{$2.3$}) discretization scheme.

\subsection{\textbf{Field dynamics $\rightarrow$ Particle-in-cell}}

A field dynamic (FD) calculation is implemented by evolving the Maxwell equations $(\textbf{1})$ and $(\textbf{2})$ in time. Such an approach under the force-free relativistic MHD approximation is preferred over the traditional standard MHD equations where plasma inertia and stress terms cannot be ignored. This makes the evolution equation easy to handle and still successfully establishes the structure of the magnetosphere. However, such an approach is incapable of handling particle motions and cannot generate acceleration mechanism. Consequently, FD calculations represents a 'first step' to establishing a complete simulation.

In the past few years, several groups (\cite{2010PASJ...62..131U}, \cite{2014ApJ...795L..22C}, \cite{2015MNRAS.448..606C}, \cite{8209414} \cite{2015ApJ...801L..19P}) have implemented particle-in-cell approaches and have gained significant insight in to a pulsar magnetosphere. These techniques not only describe the magnetic field evolution but also address the location of particle acceleration and mechanism. As a result, these simulation address a realistic magnetosphere. However, these PIC calculations still depend on the electromagnetic field solver before its motion solver is updated. The transition from a FD to PIC is obtained by replacing the current density for a force-free prescription with an equation of motion solver for particles (Boris algorithm or \cite{petri_2017}).

The eventual goal is to transition to a PIC simulation. However, this paper is mainly focused on introducing the implicit scheme (see section $2.3$) and the FD calculation for a FFE prescription. The main advantage of our approach will be the implicit update of the electromagnetic fields. The equation of motion solver for particles will be implemented in a subsequent paper.

\subsection{\textbf{Explicit $\rightarrow$ Implicit}}

Numerical modelling has played a very significant role in modern studies of the pulsar magnetosphere. Recent studies have argued for current sheets as one possible source of high-energy gamma-ray emission \cite{2014ApJ...780....3U}. In addition, non-ideal effects for high conductivity values \cite{2012ApJ...746...60L} have also been explored to achieve a realistic structure of the magnetosphere. To date, all FD and PIC numerical modelling of the magnetosphere have been carried out by the traditional \textit{explicit} finite difference approach, which is robust and has been extremely successful. 

For FD calculations the explicit nature of these code sets a limit on the time step, as the basic Courant-Fredrich-Levy (CFL) \cite{1967IBMJ...11..215C} criterion must be satisfied to avoid divergent solutions. The CFL criterion is the necessary condition required for convergence of the partial differential equation (PDE). This convergence bound and limitations on spatial and time resolution prevents a full development of the current sheets. In addition, for high conductivity magnetospheres smaller time steps are required, thereby increasing computational cost. With the current state of computational architecture, it is opportune to initiate and develop a more realistic understanding of the pulsar magnetosphere. The primary motivation for this work is thus to develop an $\textit{implicit}$ time-dependent numerical code to explore the structure of a pulsar magnetosphere. The implementation of an implicit simulation avoids the CFL limit because the update equations are $\textit{unconditionally stable}$ \cite{MOP:MOP21684}. A stable solution can be achieved by overstepping plasma frequency oscillations, thereby reducing computational cost. This will help us set up future problems, for example, the resistive problem, with less computational cost.

Similarly for PIC, implicit approach can be a very effective technique to simulate small-scale structures on a numerical grid. Traditionally, such problems would require not only small grid sizes, but also very small time steps. Because of the nature of discretization, the CFL limit for an implicit technique becomes unnecessary. Thus small-scale features within a magnetosphere can be developed more efficiently than the traditional explicit approach. Hence, an implicit or hybrid schemes can simulate small-scale features and is better equipped to achieve a self-consistent solution.

These motivations have influenced us to apply an implicit Crank-Nicolson Finite Difference Time Domain (CN-FDTD) discretization technique to our pulsar magnetosphere problem \cite{Crank1996}. For a charge-free region ($\bar \rho=0$), an important property of FDTD and Yee grid is the conservation of divergence-free ($\nabla \cdot \textbf{$\bar B$} =0$) nature of the fields and can be easily demonstrated \cite{inan_marshall_2011}.

\section{NUMERICAL METHODS AND REQUIREMENTS}

\subsection{\textbf{Computational Requirements}}

Hyperbolic Maxwell equations $\textbf{(2)}$ and $\textbf{3}$ are discretized in space and time using CN-FDTD (see \textbf{Appendix} for one such discretized update equation). These 3D linear update equations are complicated and must be solved simultaneously \cite{MOP:MOP21684}. The resulting linear equation of the type {\tt Ax = B} can be memory intensive as it involves matrices and matrix inversion. In the above linear equation, {\tt A} represent a constant coefficient matrix, {\tt x} is the unknown vector for future time step {\tt n} and {\tt B} is the known vector for the previous time step {\tt n-1}.

\begin{figure}[tbp]
\centering
\includegraphics[width=\textwidth]{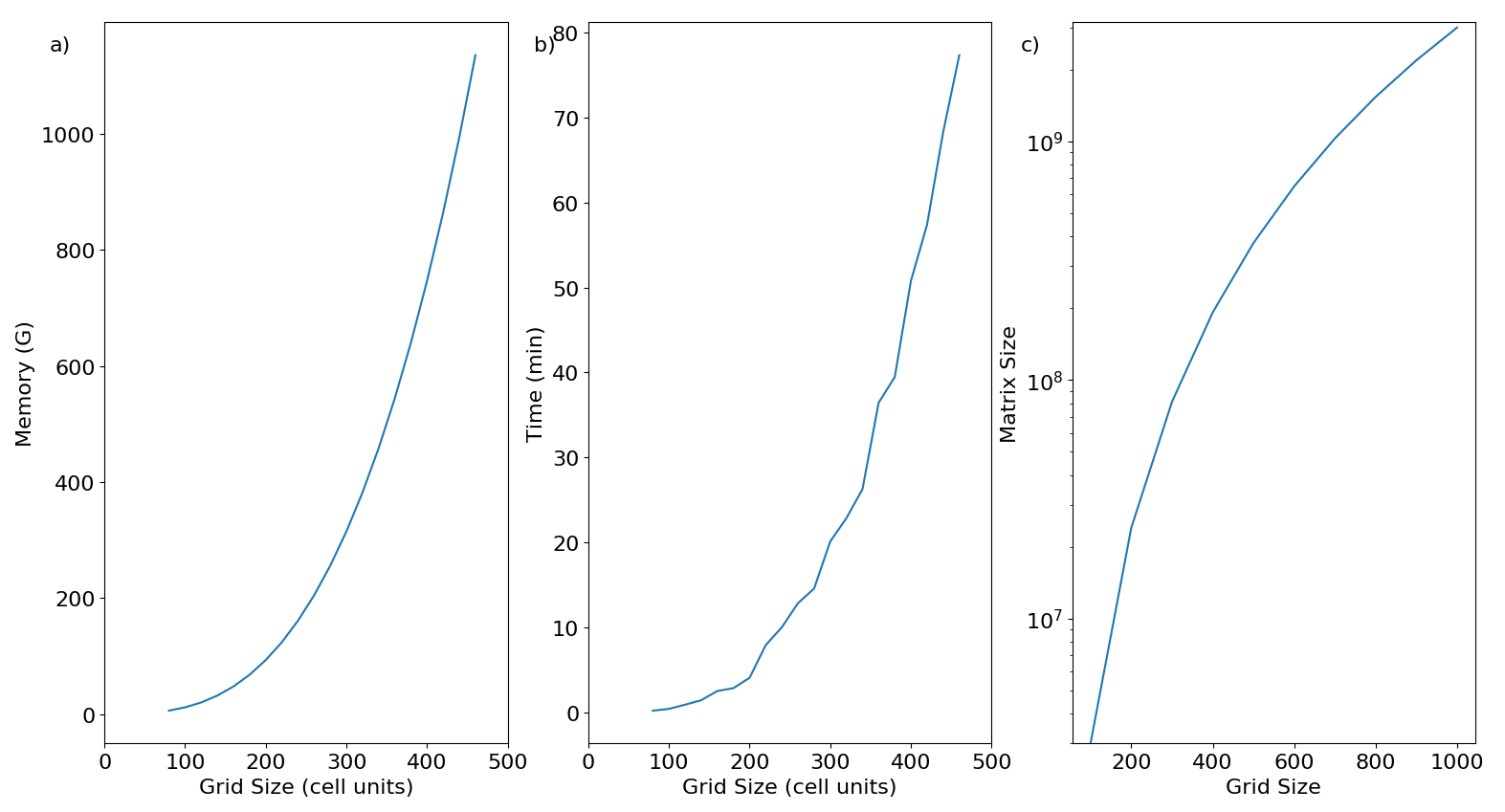}
\caption{Computational requirements as a function of grid size.}
\label{fig:1}       
\end{figure}

For each time step, the time and computational memory required to solve our implicit equations as a function of grid size are shown in figure \textbf{1a} and \textbf{1b} respectively. The size of the matrix (MS) as a function of grid size can be represented by (see figure \textbf{1c})

\begin{equation}
MS = \alpha*\beta^D
\end{equation}

Here, $\alpha$ is number of unknown vector equations, $\beta$ is the range along one grid dimension and D is number of grid dimensions.

A proper choice of linear equation solver thus becomes extremely critical. After much deliberation\footnote{http://www.netlib.org/utk/people/JackDongarra/la-sw.html}, we chose to implement the Library of Iterative Solvers (Lis)\footnote{http://www.ssisc.org/lis/. Free open ware software, subroutines to handle extremely large matrices and support for large number of iterative solvers are some of the advantages of Lis} to solve our system of linear equations. This open-source solver is not only easy to use but also handles large  sparse matrices in an efficient matrix market (MM) format. Finally, instead of central symmetry \cite{2009A&A...496..495K} we have implemented a complete $360^o$ evolution of fields.

\subsection{\textbf{Electrodynamic Requirements}}

The main objective of this work is an implicit FD calculation under FFE approximation for a pulsar magnetosphere. Our aim is to reproduce the CKF type magnetosphere which is a new benchmark in pulsar modelling. The prescription for current density under FFE approximation is given below (equation $\textbf{8}$), where the terms have their usual meaning. Here, term 1 represents perpendicular component of the current density with drift velocity $\bar E \times \bar B$ and term 2 represents the current parallel to fieldlines. Due to complexities involved in setting up the matrices for term 2 of the current density this term is ignored for the present work. Such a measure however, warrants enforcing the ideal MHD condition (see section $3.2.2$).

\begin{multline}
\qquad \qquad \qquad \bar J = \frac{c}{4\pi} \nabla \cdot \bar E \frac{\bar E \times \bar B}{B^2} \\
+ \frac{c}{4\pi}\left [\frac{\bar B\cdot\nabla\times\bar B-\bar E\cdot\nabla\times\bar E}{B^2}\right]\bar B \qquad \qquad
\end{multline}

On a Cartesian grid, the electric and magnetic fields are initialized to zero and a pure dipole respectively. The update vectors are then passed through an iterative solver from Lis resulting in the evolution of the fields. However, the final update vector must undergo the following tests to avoid any nonphysical conditions.

\subsubsection{Subluminal validity}

The drift velocity term i.e.$\left(\frac{\bar E \times \bar B}{B^2}\right)$, which arises from the perpendicular component of the current density must be maintained at subluminal values. As a result, for every time step, the $\left|E\right| \leq \left|B\right|$ condition must be controlled for each component. For regions where subluminal conditions are violated i.e. $\left|E\right| > \left|B\right|$, a normalization factor $\frac{\left|\bar B\right|}{\left|\bar E\right|}$ is obtained, which is then multiplied with the corresponding $\left|\bar E\right|$ component.   

\subsubsection{Enforcing ideal MHD conditions}

The Maxwell equations on their own are incapable of enforcing the ideal MHD condition $\bar E \cdot \bar B=0$. Hence, for each time step, we interpolate the fields to the same location and then calculate the change in field components:
\begin{equation}
\Delta E_{i,j,k} = \frac{\bar E \cdot \bar B}{\left |B\right |} B_{i,j,k}
\end{equation}
The final update vector is rewritten as
\begin{equation}
E_{i,j,k} = E_{i,j,k} - \Delta E_{i,j,k}
\end{equation}
This validation is a prerequisite as only the perpendicular component of the current density is presently considered. If the entire current density term (equation $8$) is implemented, this enforcement will be unnecessary.

\subsubsection{Rotational effects}

The electric and magnetic fields are updated throughout the computational domain. For a smooth transition of fields from interior to exterior of the star we using a smoothing kernel and is done as follows. The update equations along with the corotating electric field $ \left(\bar E=\bar \Omega \times r \times \bar B \right)$ and coordinate-free magnetic dipole field $\left(\bar B=\frac{3\hat{r}\bar m(t)\cdot \hat{r}-\bar m(t)}{r^3}\right)$ are linked with each other by a smoothing kernel \cite{2006ApJ...648L..51S};
\begin{equation}
    f(i,j,k) = f(i,j,k) + \left [h(i,j,k) - f(i,j,k) \right] * s(i,j,k)
\end{equation}
Here, $\Omega$ is the angular rotation of the star, $\bar m(t)$ is the dipole moment depending on the inclination angle between rotation axis and dipole moment,$f(i,j,k)$ is the localized field component obtained from discretization of equation (4) and (5). Finally, $h(i,j,k)$ is the co-rotating and rotating dipole fields for $\bar E$ and $\bar B$ respectively. A hyperbolic smoothing function $(s)$ is used to prevent discontinuities in field lines.

\begin{figure}[tbp]
  \centering
  \includegraphics[width=0.7\textwidth]{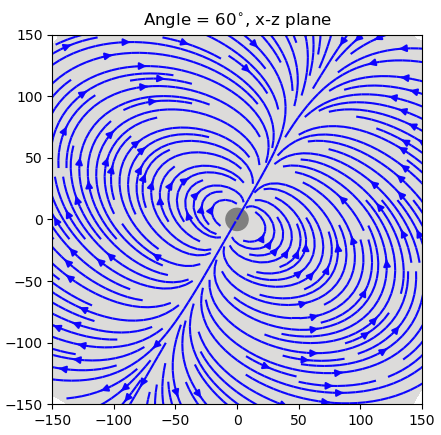}
\caption{Initialized dipole field for an inclined $60^o$ rotator surrounded by vacuum.}
\label{1.9}       
\end{figure}

\section{\textbf{PROBLEM SET UP}}

\subsection{Vacuum solution}

For a vacuum solution, there are no charges surrounding the NS and it can be reproduced in a FD calculation by neglecting the source term in equation $(3)$. The magnetic field for an inclined rotator is initialized to a dipole (figure \textbf{2}), thereafter the implicit update equations for $\bar J = 0$ are evolved in time. Except for the rotational effects of section $3.2.3$ none of the electrodynamic requirements are required to be verified for a vacuum solution.

\begin{figure}[!tbp]
\centering
\includegraphics[width=\textwidth]{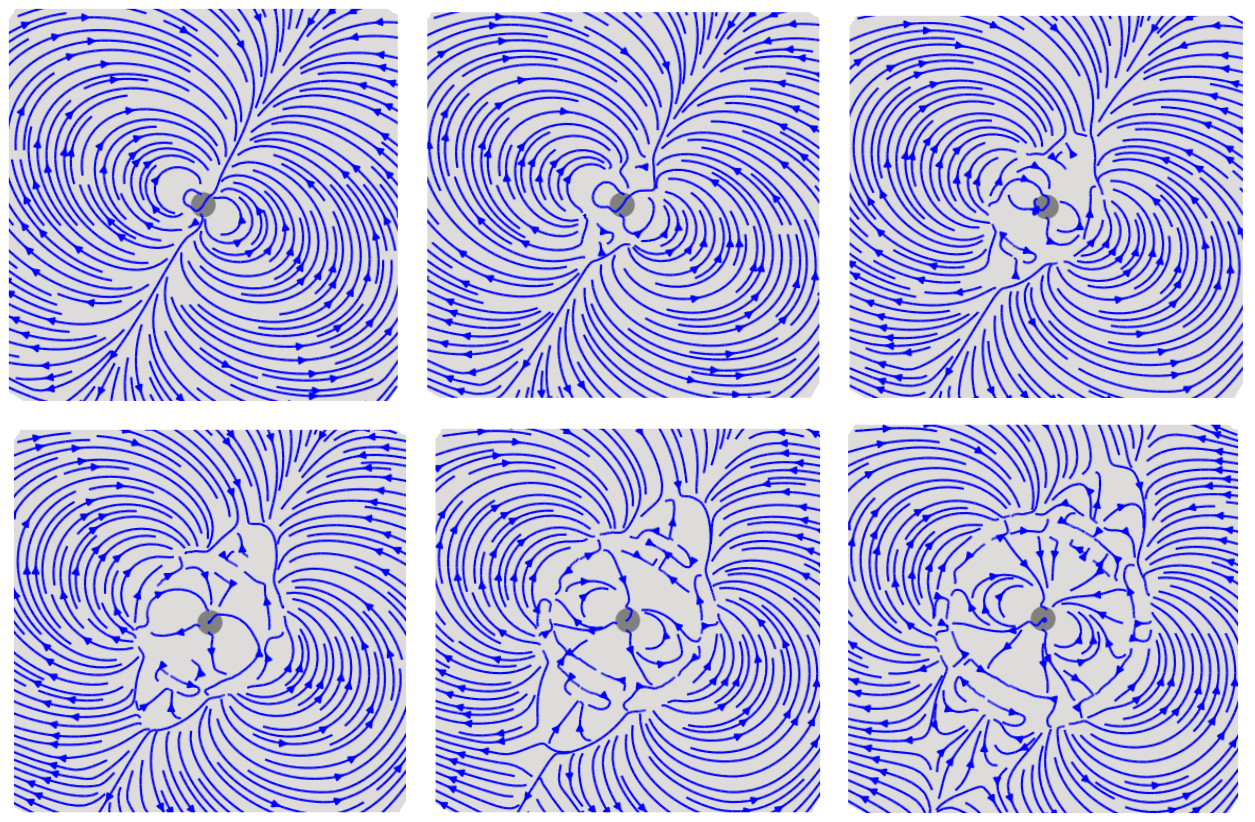}
\includegraphics[width=\textwidth]{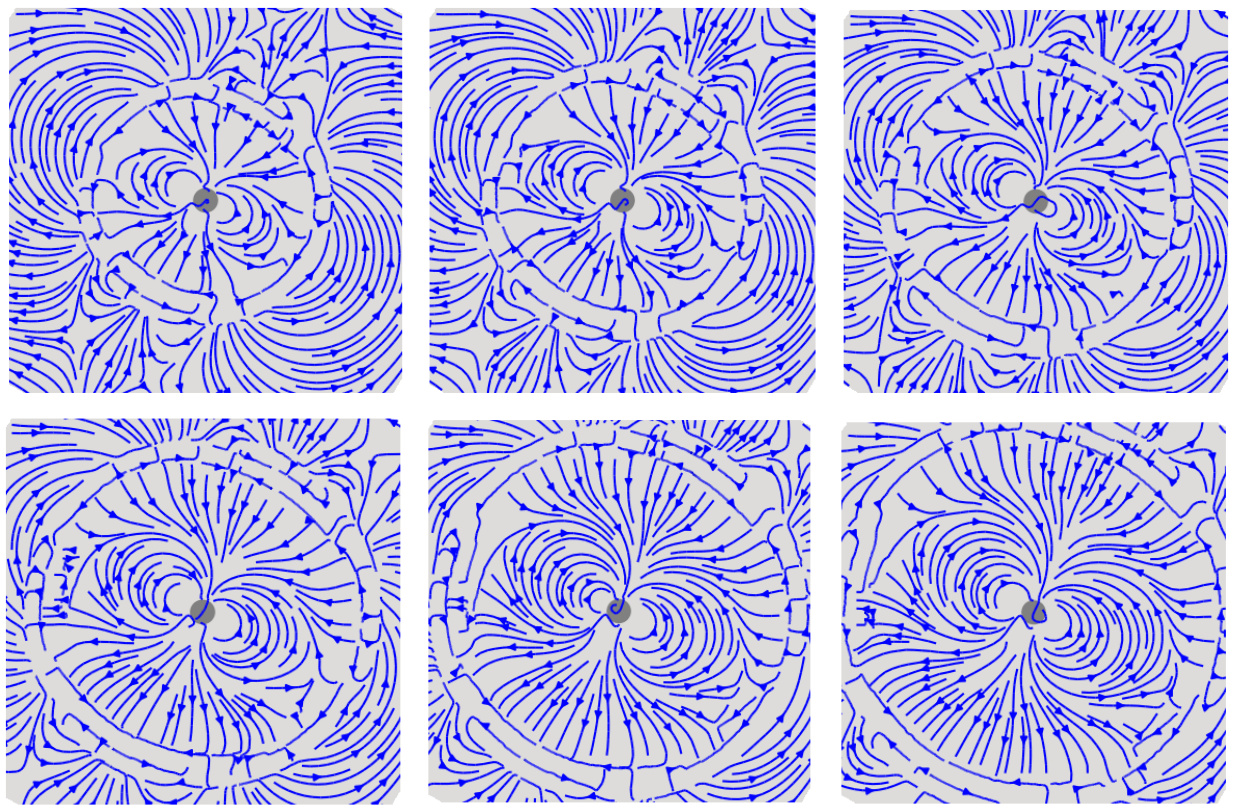}
\caption{Time evolution for a $60^o$ inclined dipole rotator (x-z plane) in vacuum. Each frame is $\approx 1/12$ of a rotation.}
\end{figure}

For $t>0$, an electromagnetic wave is radiated in the outward direction till it reaches the computational boundary in $\approx 3/4$ of a rotation. As the transient wave passes through the grid, field lines return back to the star and the result mimics the analytical solution of a dipole field in vacuum. Figure \textbf{3} (plotted in x-z plane) shows a time sequence for one such inclined rotator. The transient wave crosses the light cylinder in $\approx 1/6$ of the rotation period and thus demonstrates that the wave is not travelling at speed greater than the speed of light. The vacuum solution was also verified for several other inclination angles.

\begin{figure}[!tbp]
\centering
\includegraphics[width=\textwidth]{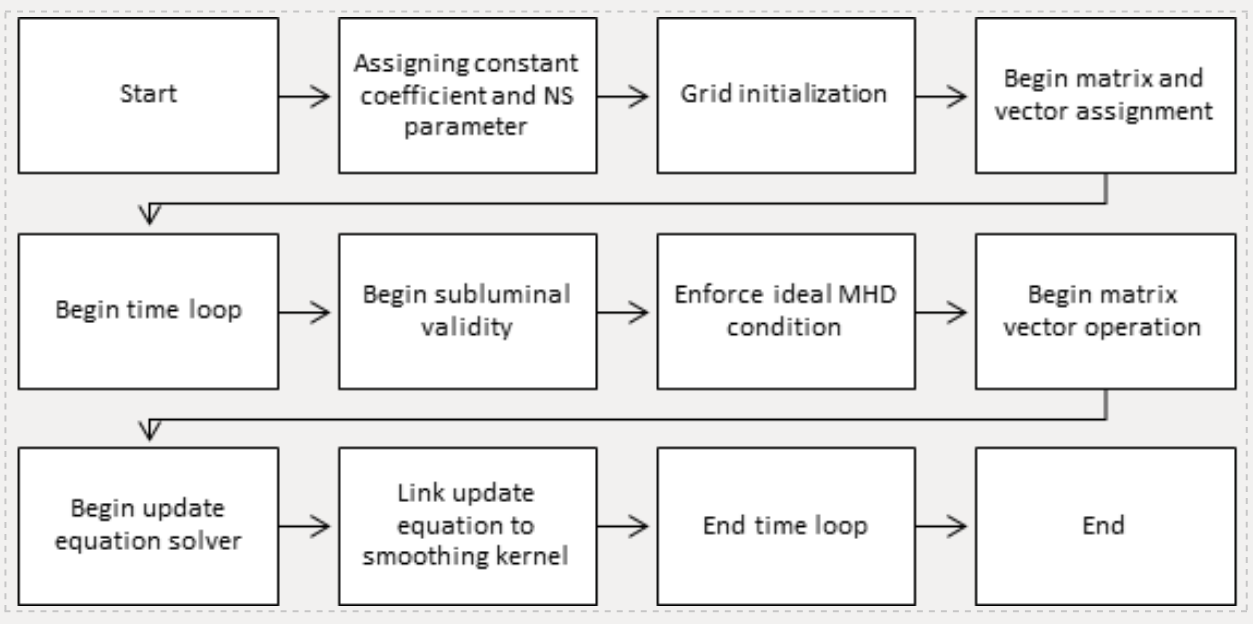}
\caption{Computational flowchart for the implicit update equations to achieve a steady state}
\label{fig:4}       
\end{figure}

\subsection{FFE solution}

The vacuum solution provides a necessary test of our implicit code. But the FFE is the first step toward a realistic solution. Using the complete Maxwell equation ($\textbf{2}$ and $\textbf{3}$) and only the perpendicular component of the current density (term 1 of equation $\textbf{8}$), update equations are developed (See appendix for one such update). Other flavors of current density (\cite{2007ApJ...667L..69G}, \cite{2012ApJ...746...60L}) will be evolved for a later paper. These equations are generated under ideal MHD and FFE (equation $\textbf{4}$ and $\textbf{5}$) conditions.  Once the update equations are generated, the problem at hand is "relatively" straight forward to set up and is done as follows:

\begin{itemize}
    \item For an inclined dipole rotator of inclination angle $\psi$, a computational grid is generated by initializing the electric and magnetic fields to zero and a pure dipole respectively.
    \item Using the Lis solver, every constant coefficient matrix (like matrix $\tt{A}$ in $\tt{Ax=b}$ linear equation and other matrices used in boundary conditions) and vectors necessary for the update equations are generated.
    \item Within the time evolution loop, we first implement the time-dependent dipole moment and corotating electric field terms and set the star in rotation. 
    \item The final update vector is checked for \textit{subluminal validity}, \textit{ enforcement of the MHD condition} and \textit{rotational effects}.
    \item The update vector is then passed to an iterative solver available within the Lis libraries, to update the field vector for the future time step. After several rotations the star settles into a stationary solution. 

\end{itemize}
    
Refer to figure \textbf{4} for a complete computational flowchart.

\section {RESULTS}

For a 3-D Cartesian grid size of $200^3$, the implicitly discretized Maxwell equations were set in a time evolution loop. The radius of the neutron star was set at 9 cells and the light cylinder radius $(R_{LC}=c/\Omega)$ was calculated to form at 29 cells from the neutron star (in the rotation axis frame). The magnetic field is initialized to a dipole (figure \textbf{6 top}). After approximately $1/4$ of a stellar rotation, the field settles and evolves steadily throughout the grid. However, some visible oscillations in the luminosity flux are still persistent as seen by Kalapotharakos \cite{2009A&A...496..495K}. These oscillations seem to to be an artifact of our cartesian grid as it settles down after $1/2$  rotations. A steady state for the fields is achieved after 1.75 rotations, thereafter we have closed field lines that extend outward to the light-cylinder and open field lines that extend beyond this characteristic length. The toroidal fields beyond the light cylinder causes the fields to \textit{sweep back} in the direction opposite to the rotation of the central star. A steady state CKF- type magnetosphere has thus been achieved using the implicit discretization approach.

\begin{figure}[!tbp]
\centering
\includegraphics[width=\textwidth]{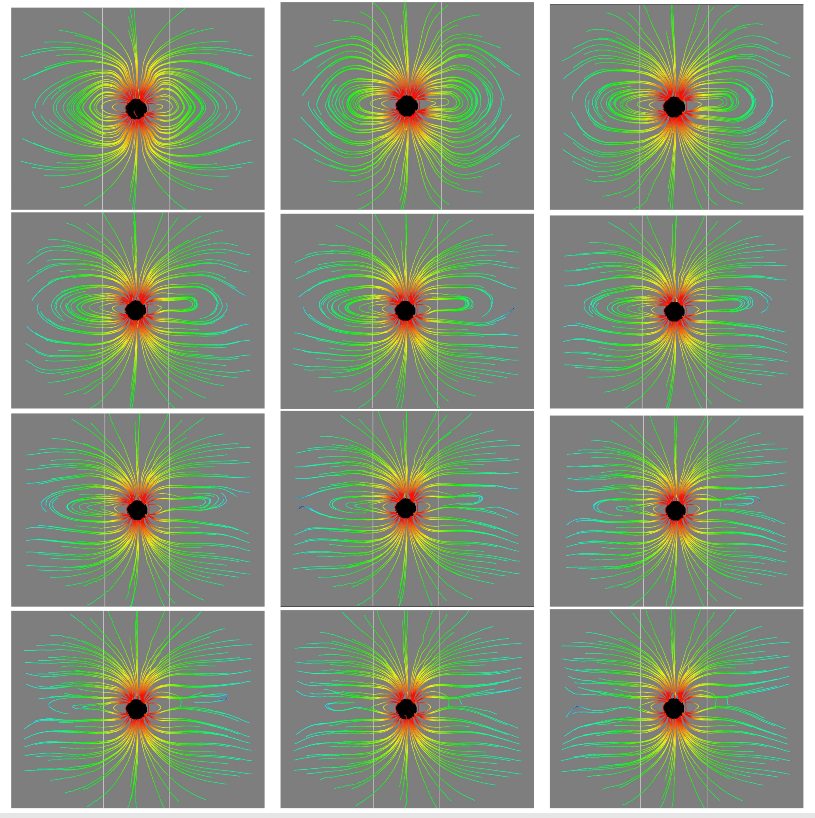}
\caption{Time sequence  of the magnetic field for a $60^0$ oblique rotator. The initial dipole field at t=0 achieves a steady state in less than two rotations. Each frame is after $1/6^{th}$ of a rotation. Vertical lines represent the light cylinder. Color represents field strength}
\label{fig:5}       
\end{figure}

A time sequence of the magnetic field achieving this steady state is shown in figure \textbf{5}. This implicit code was run for approximately 6 stellar rotations and the steady state conditions were maintained throughout the run time. Our code was also able to capture plasmoid ejection near the light cylinder (Figure \textbf{6} bottom), which seems to be essential to maintain the CKF type magnetosphere. These effects have been observed by previous pulsar modelling techniques and verified implicitly by our group. These effects can therefore, be attributed as a real magnetospheric phenomenon.

\begin{figure}[!tbp]
\centering
\includegraphics[width=\textwidth]{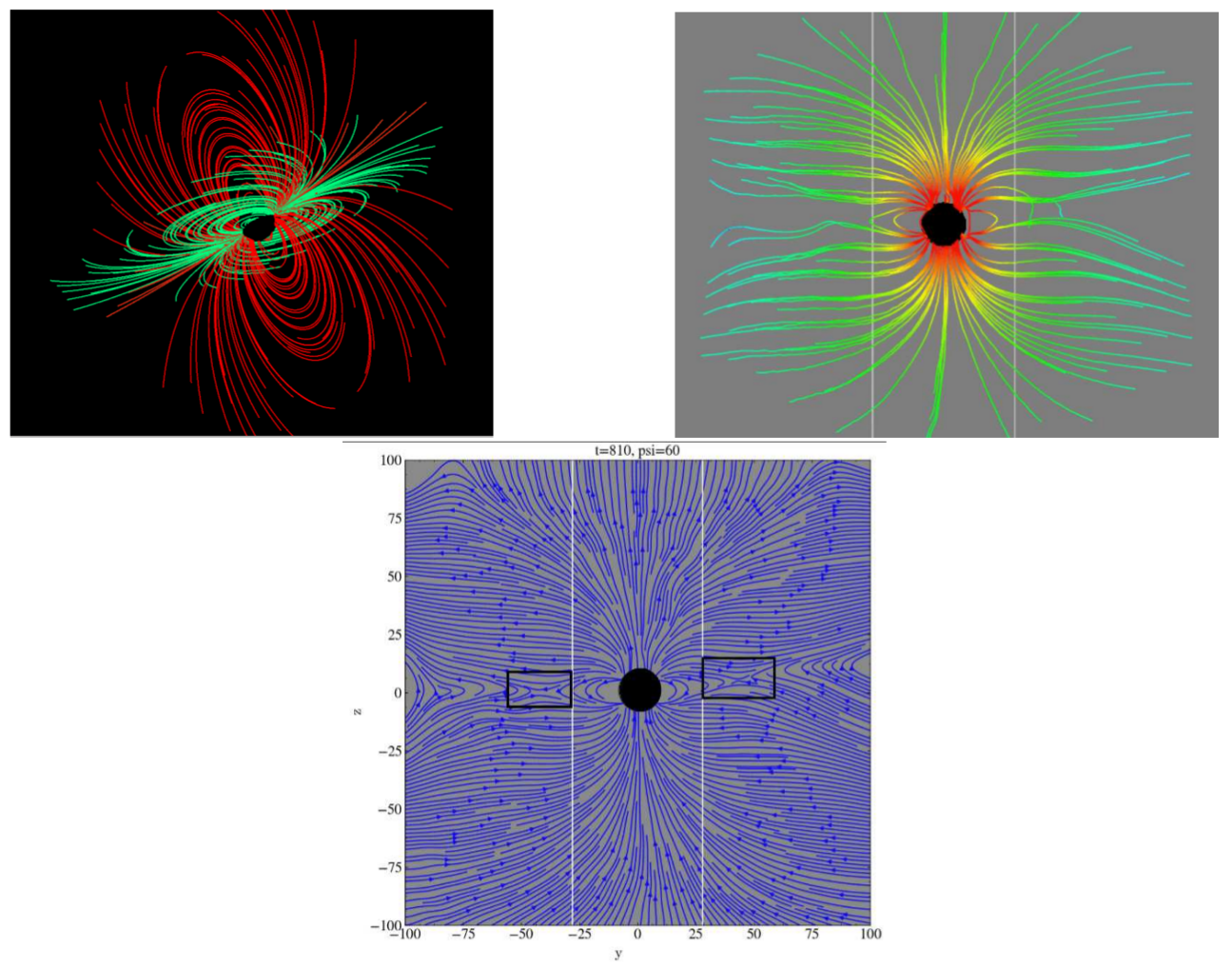}
\caption{\textbf{top left)} Initialized dipole field for a rotator inclined at $60^0$. \textbf{top right)} Evolution of the field after steady state is achieved. Vertical lines represents light cylinder and color represents field strength. 
Refer to fig. \textbf{4.4} for a time sequence of the above evolution. \textbf{bottom)} Same as above but with dense field lines to represent generation of equatorial reconnection and plasmoid events (black rectangles)}
\label{fig:6}       
\end{figure}

\begin{figure}[!tbp]
\centering
\includegraphics[width=\textwidth]{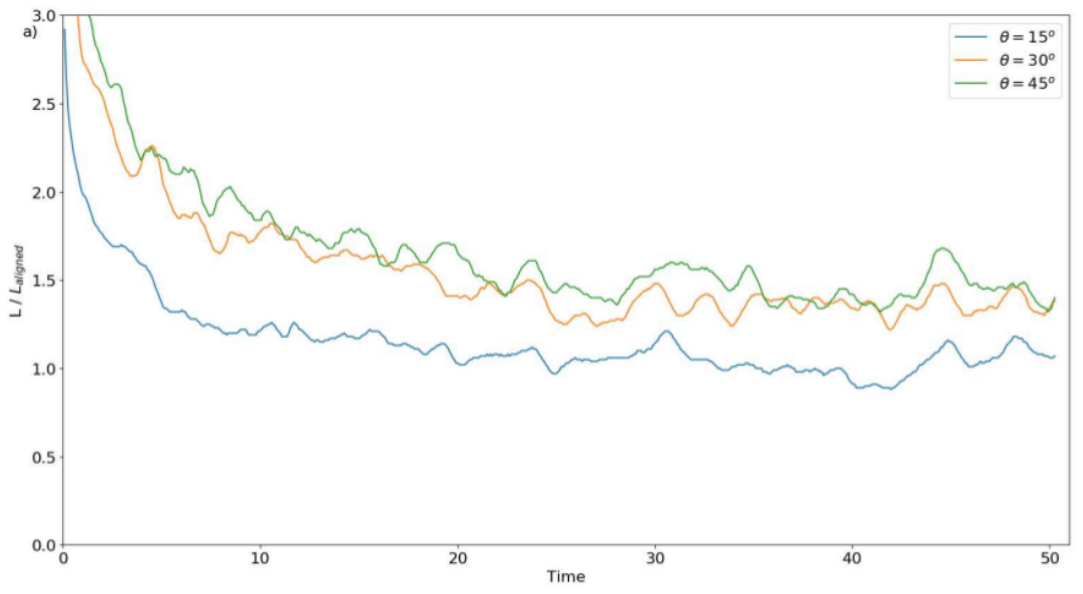}
\includegraphics[width=\textwidth]{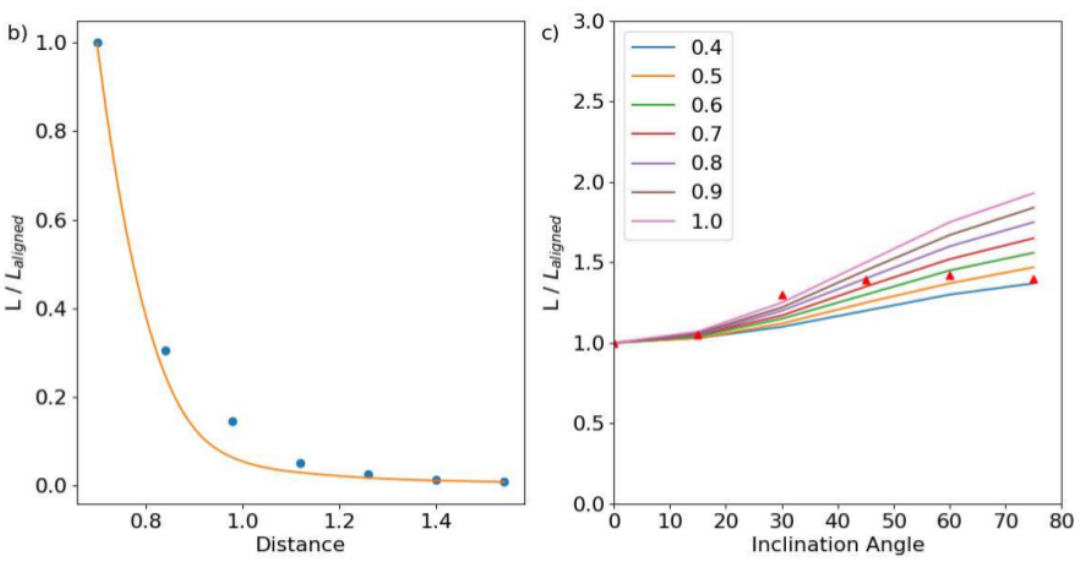}
\caption{a) Poynting flux as a function of time (in the units of $r_{LC}/c$) for various inclination angles. b) Flux as a function of distance (in the units of $r_{LC})$  for a ${60^0}$ rotator. Inverse square law is fit (solid line) to simulation data (dots) c) Luminosity as a function of inclination angle and comparison of our simulation data (triangles) for various coefficient value in equation 12.}
\label{fig:7}       
\end{figure}

The stability of our luminosity values over several stellar rotations can be seen in figure \textbf{7a}. From the luminosity evolution curve it can be concluded that our implicit scheme is fairly stable for several stellar rotations. Within two stellar rotations the magnetic fields settle into a steady state. The inverse square relationship of flux as a function of distance can also be seen in figure \textbf{7b}. 

An important study in pulsar magnetospheres is to understand the luminosity evolution for various inclination angles. From our implicit update equations, luminosity can be computed by calculating the Poynting flux $(\bar S=\bar E \times \bar B)$ around a cube centered on the neutron star. Figure \textbf{7c}, represents the Poynting flux as a function of inclination angle for various coefficient of the inclination angle. Similar to equation \textbf{1} and \textbf{6}, the best fit for the luminosity as a function of inclination angle is given by: 

    \begin{equation}
        L \approx \frac{\mu^2\Omega^4}{c^3}\left (1+0.6\,sin^2 \theta \right )
    \end{equation}

Our result matches (within $5\%$) for low inclination angles. However, for higher angles our results are consistent within $15\%-20\%$, which might be a result of numerical dissipation within our grid.

Our implicit code was able to run successfully even without an absorbing boundary like a Perfectly Matched Layer (PML). We were able to run our simulation for a longer time without the reflected wave affecting our physical grid. We will be implementing PML for an efficient open boundary for all future work. But, for a smaller grid size, our preliminary results qualitatively agrees with the results of Spitkovsky \cite{2006ApJ...648L..51S} and Kalapotharakos \cite{2009A&A...496..495K}.

We have successfully reproduced the CKF-type magnetosphere,which is now a widely-accepted structure of a pulsar magnetosphere. We were also successful in  reproducing the solution for a vacuum model. The structure of the magnetosphere was achieved  by implementing an unconditionally stable CN-FDTD implicit scheme. We have therefore demonstrated that the force-free solution differs from the vacuum solution such that the field lines beyond the light cylinder do not return to the star. Our implicit scheme now satisfies the two extreme cases of a pulsar magnetosphere: the \textit{vacuum} and \textit{force-free} approximations.

\section{FUTURE WORK}
This paper is the first of a small series exploring a pulsar magnetosphere. Here, we have described our code and used it to independently verify the structure of the magnetosphere using our implicit technique. Our next steps includes:

\begin{enumerate}
    \item To implement the implicit scheme for a resistive pulsar magnetosphere \cite{2012ApJ...746...60L}. Traditional techniques for resistive problems have a drawback as the basic CFL criteria must be satisfied. This makes explicit schemes computationally expensive.  An implicit code with proper choice of spatial and temporal resolution can be very effective, thereby reducing computational cost. This would be the biggest advantage of using an implicit scheme.
    \item To implement an implicit discretization scheme to a particle-in-cell (PIC) modelling. Several PIC simulation of pulsar magnetosphere have been performed \cite{2014ApJ...785L..33P}, \cite{2015ApJ...801L..19P} and modelling  high-energy light curves using PIC have also been developed \cite{2016MNRAS.457.2401C}. However, these codes are unable to resolve small scale structure to understand the microphysics of, for example, pair creation \cite{2017arXiv171202409H}. If implicit codes can be developed with better resolutions, we expect to simulate these sites very effectively.
    \item To implement efficient absorbing boundaries. Our initial attempt at the structure of the magnetosphere was developed without an efficient lossy material surrounding the FDTD grid (similar to perfect electric conductor, PEC=0). However, to implement realistic magnetospheric effects an effective absorbing boundary is essential. Currently the best absorbing boundaries for waves approaching the boundaries at large angles is the \textit{perfectly matched layer} (PML).
    \item To develop a global structure would be the ultimate goal of our project. These global structure  models should reproduce the observed light curve phenomenology available in the Second Fermi Pulsar Catalog \cite{2013ApJS..208...17A}. This can be achieved by different form of current density terms and incorporating it into the update equations. By developing various prescription for an non-ideal MHD condition, we can therefore establish a range of solutions from vacuum to force-free \cite{2012ApJ...746...60L}, \cite{2014ApJ...793...97K}. 
    \item The current implicit equations are maximally optimized for a state-of-the-art computational architecture without parallelization. For a dynamic programming algorithm like ours implementing MPI or OpenMP can minimize some of the computational costs, especially for problems where large grid size is required. However, this would require changes to the source code of the Lis solver to make it compatible for parallelization.
\end{enumerate}

Some long term projects that can be achieved using the implicit schemes will be to explore a system of binary neutron stars. Using the implicit scheme we can examine the interacting magnetosphere and its effects on Pyonting flux and field configuration. However, this would require the implementation of Adaptive Mesh Refinement (AMR) into the code as the interacting grids would be significantly larger than what is currently implemented. Another long term project that we would like to explore is the magnetosphere for a magnetar. Using approximation in quantum force-free electrodynamics (Q-FFE) we can verify whether the magnetosphere of a magnetar deviates from classical results \cite{2016JCAP...05..042F}.

\section{APPENDIX}
\subsection{\textbf{Appendix A}}
The general form of the CN discretization scheme is given as (Crank 1947)
\begin{equation}
\frac{u_i{^{n+1}}-u_i^{n}}{\Delta t} = \frac{1}{2} \left[F_i^{n+1} \left(u,x,t,\frac{\partial u}{\partial x},\frac{\partial ^2 u}{\partial x^2} \right) + F_i^n \left(u,x,t,\frac{\partial u}{\partial x},\frac{\partial ^2 u}{\partial x^2} \right)\right]
\end{equation} Applying the above expression to a Maxwell equation in scalar form gives (Yang 2006) \begin{multline}
\qquad \qquad \qquad \qquad \qquad \qquad \qquad
E_z^{n+1}(i,j,k) - \\ \frac{c\Delta t}{2} \left[\frac{B_y^{n+1}(i+1,j,k)-B_y^{n+1}(i,j,k)}{\Delta x} - \frac{B_x^{n+1}(i,j+1,k)-B_x^{n+1}(i,j,k)}{\Delta x} \right] \\
= \frac{c\Delta t}{2} \left[\frac{B_y^{n}(i+1,j,k)-B_y^{n}(i,j,k)}{\Delta x} - \frac{B_x^{n}(i,j+1,k)-B_x^{n}(i,j,k)}{\Delta x} \right]
\end{multline}

Repeating the steps for the other equations and modifying them to get the update equation of the form

\begin{multline}
\left [1+q_{11}+q_{12} \right] E_z^{n+1}(i,j,k)-\frac{q_{11}}{2} \left [E_z^{n+1}(i,j-1,k)+E_z^{n+1}(i,j+1,k) \right ] \\ - \frac{q_{12}}{2} \left [E_z^{n+1}(i+1,j,k)+E_z^{n+1}(i-1,j,k) \right ]  \\ + q_{13} \left [E_y^{n+1}(i,j,k-1)-E_y^{n+1}(i,j+1,k-1)-E_y^{n+1}(i,j,k)+E_y^{n+1}(i,j+1,k) \right] \\ + q_{5} \left [E_x^{n+1}(i,j,k-1)-E_x^{n+1}(i+1,j,k-1)-E_x^{n+1}{i,j,k}+E_x^{n+1}(i+1,j,k) \right] \\ = \left [1-q_{11}-q_{12} \right] E_z^n(i,j,k) +\frac{q_{11}}{2} \left [E_z^n(i,j-1,k)+E_z^n(i,j+1,k) \right ] \\ + \frac{q_{12}}{2} \left [E_z^n(i+1,j,k)+E_z^n(i-1,j,k) \right ]  \\ - q_{13} \left [E_y^n(i,j,k-1)-E_y^n(i,j+1,k-1)-E_y^n(i,j,k)+E_y^n(i,j+1,k) \right] \\ - q_{5} \left [E_x^n(i,j,k-1)-E_x^n(i+1,j,k-1)-E_x^n{i,j,k}+E_x^n(i+1,j,k) \right] \\ + q_{14} \left[B_y^n(i+1,j,k) -B_y^n(i,j,k)\right ] - q_{15} \left[B_x^n(i,j+1,k) -B_x^n(i,j,k)\right ] \\ - 4\pi \Delta t J_z^n(i,j,k) \qquad \qquad \qquad \qquad \qquad \end{multline}
Here, \begin{align*}
q_{11}=\frac{c^2\Delta t^2}{2\Delta y^2} & & q_{12}=\frac{c^2\Delta t^2}{2\Delta x^2} & &  q_{13}=\frac{c^2\Delta t^2}{4\Delta y \Delta z}\\
q_{3}=\frac{c^2\Delta t^2}{4\Delta x \Delta z} & & q_{14}=\frac{c\Delta t}{\Delta x} & & q_{15}=\frac{c\Delta t}{\Delta x}
\end{align*}
Update equations for other vector equations are obtained in a similar manner.

\begin{figure}[!tbp]
  \begin{center}
    \includegraphics[width=0.5\textwidth]{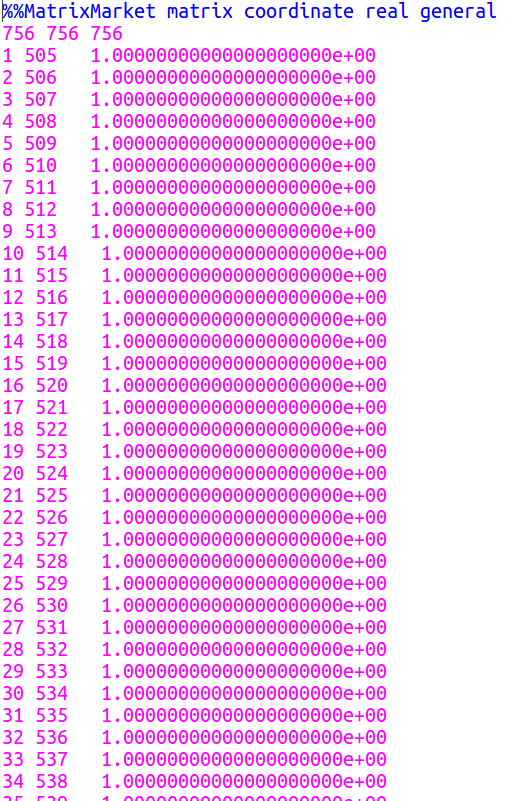}
  \end{center}
  \caption{MM FILE}
\end{figure}

\subsection{\textbf{Appendix B}} Matrix Market (MM) Format is an extremely efficient approach to represent sparse matrices. This is because only the non-zero elements of the matrix are stored, reducing the size of the data file. For our pulsar magnetosphere problem, the structure of the matrix is extremely symmetric above and below the main diagonal. But general, skew-symmetric and Hermitian matrices can also be adequately represented in MM format. A simple example of a file in MM format is shown in figure \textbf{8}. The first line indicates that the file is in MM \textit{coordinate} format and values are represented in \textit{real} and \textit{general} form. The second line of the file indicates the number of rows, columns and number of non-zero elements in the matrix. This is followed by the matrix data. The non-zero elements of the matrix are then represented by the row and column number followed by the value assigned to that element.
\newline
\newline
\textbf{Declarations} 
\newline
\newline
\textbf{Availability of data and material} 
This paper is part of the dissertation presented by Sushilkumar Sreekumar in partial fulfillment of the requirements for the degree of Doctor of Philosophy in Physics at University of Texas San Antonio. The simulation code developed for this work will made available after the submission of the dissertation.
\newline
\newline
\textbf{Authors’ contributions} 
The implementation of the CN-FDTD to a pulsar magnetosphere and all of the tests were performed by SS. The project was initiated and closely supervised by EMS. The results and contribution to the final manuscript were performed by SS and EMS.
\newline
\newline
\textbf{Competing interests} 
The authors declare that they have no competing interests. 
\newline
\newline
\textbf{Acknowledgments} 
This work received computational support from UTSA's HPC cluster Shamu, operated by the Office of Information Technology. We would like to thank Anatoly Spitkovsky for his scientific input during the development of the code. We would like to thank Akira Nishida for his computational input in optimizing our implicit code. Finally, we would also like to thank the Vaughan Family Endowment for their partial support.
\newline
\newline
\textbf{Authors' information}\\ 
Sushilkumar Sreekumar \\
Department of Physics and Astronomy\\
University of Texas San Antonio (UTSA) \\ \\
Eric M Schlegel \\
Vaughan Family Professor\\
Department of Physics and Astronomy \\
University of Texas San Antonio (UTSA)

\bibliographystyle{plain}
\bibliography{123.bbl}

\end{document}